# Negative differential resistance: another banana?

J. Li,[1] H.-F. Zhang,[1] G.-Q. Shao,[1,a)] B.-L. Wu,[2] and S.-X. Ouyang[3]

[1]*State Key Laboratory of Advanced Technology for Materials Synthesis and Processing, Wuhan University of Technology, Wuhan, 430070, China*

[2]*Key Laboratory of New Processing Technology for Nonferrous Metals and Materials, Guilin University of Technology, Guilin, 541004, China*

[3]*China Building Materials Academy, Beijing, 100024, China*

Just like the artefact found in ferroelectric hysteresis loops, the nearly identical NDR effect shown in $Sr_3Co_2Fe_{24}O_{41}$, $TiO_2$, $Al_2O_3$, glass and even banana skins is confirmed to be a kind of water behavior. The combination of water induced tunneling effect, water decomposition and absorption plays a crucial role in the NDR effect. The results and mechanism demonstrated here illustrate that much attention should be paid to the chemical environment when studying electrical properties of materials / devices.

**I. INTRODUCTION**

Non-ferroelectrics (even bananas) could exhibit closed loops of switched charge versus applied voltage due to the leakage current[1]. Current-voltage (*I* - *V*) hysteresis loops, representing a negative differential resistance (NDR) effect, has not been reported to date in non-NDR materials / devices, whatever they are ferroelectrics or not.

The NDR effect have attracted considerable attention due to the potential applications in next-generation resistive-switching random access memories (RRAM)[2-6] and missing memresistors[7]. They are related to a combination of thermal, chemical and electronic / electrostatic effects[4] and could be caused by charge storage[8], electrically field-induced Mott-transition[2, 9], filamentary conducting[2, 9], oxygen-vacancies migration[10-12], polarization relaxation[13], ferroelectric polarization-reversal[14], space-charge-limited currents (SCLC)[15] and magnetoresistive dielectric-breakdown[16] *et al.*. In many studies, a metal-insulator-metal (MIM) configuration is adopted because of its simple structure and compatibility with the fabrication process of conventional complementary metal-oxide-semiconductor (MOS) field-effect transistors[17].

Some researchers observed unusual electrical effect that was induced by chemical environment, *e.g.* ambient moisture. Kim *et al*.[18] and Aguirre *et al*.[19] found that carbon nanotube field-effect transistors demonstrate *I* - *V* hysteresis due to charge trapping by water molecules[18]. Chakrapani *et al*. suggested

---

a) Author to whom correspondence should be addressed. Electronic mail: gqshao@whut.edu.cn.

that charge transfer in adsorbed water should be considered when interpreting the substantial conductivity of undoped diamond exposed to air[20]. For AlGaN / GaN high-electron-mobility transistors (HEMTs) under ambient moisture, Gao *et al.* found current collapse by the ionization and deionization of water molecules at the device surface[21, 22].

In brief, it has been accepted that water molecules can significantly affect the electrical properties of materials / devices, but the above interpretation for NDR effect are all based on intrinsic properties with water-strengthened bulk / interface / surface effects for materials / devices. Water can coat all hydrophilic surfaces under ambient conditions, and the first water adlayers on a solid often dominate the surface behavior[23]. Charge transfer related to water is indeed an important phenomenon that is always overlooked because the consequences are easily confused with the intrinsic properties of materials[19]. Is it possible the NDR effect existing in non-NDR materials / devices? Or is the NDR effect independent of materials / devices under a different chemical environment?

In this work, the symmetrical NDR effect is determined in $Sr_3Co_2Fe_{24}O_{41}$ ($Sr_3Co_2Z$), $TiO_2$, $Al_2O_3$, glass and banana skins. The Z-type hexaferrite $Sr_3Co_2Z$ has been studied for its magnetoelectric and magnetodielectric properties[24-26] in which a NDR effect is expected. The $TiO_2$ and $Al_2O_3$ are intrinsic NDR materials[27-30], while the glass (not $SiO_2$ single crystal) and banana skins are not. A model is proposed to interpret the observed phenomena.

## II. EXPERIMENT

The method for preparing $Sr_3Co_2Z$ was reported elsewhere[26, 31]. Before measurement of electrical properties, Ag electrodes were pasted on polished plates ($\Phi 11.72 \times 0.84$ mm) and fired in oxygen at 830 °C for 10 min. The current density as a function of electric field ($\sigma$ - $E$ or $I$ - $V$) was measured by Keithley 2410 Source Meter (Keithley Instruments, Inc., USA). The voltage was swept in a sequence of $0 \rightarrow U_{max} \rightarrow 0 \rightarrow -U_{max} \rightarrow 0$. The polarization - electric field ($P$ - $E$) hysteresis loops were tested by a standard ferroelectric testing system (Premier II, Radiant Technologies, Inc., USA) at 100 Hz. The impedance spectroscopy was measured by Precision Impedance Analyzer 6500B (Wayne Kerr Electronics Inc., Britain) from 100 Hz to 1 MHz. All measurements were conducted under room temperature.

## III. RESULTS AND DISCUSSION

### A. *σ* - *E* and *P* - *E* curves

Fig. 1 shows current density ($\sigma$) and polarization ($P$) as a function of electric field ($E$) for $Sr_3Co_2Z$ in various environment. The dry, ambient and moist environments represent the relative humidity of 0 ~ 10 %, 50 ~ 70 % and 80 ~ 100 %, respectively. The $Sr_3Co_2Z$ in moist and ambient environments with configuration i exhibit a symmetrical NDR effect (curves 1 ~ 3). At a low electric field, $\sigma$ - $E$ curve exhibits a linear feature. As the applied field increases, the $\sigma$ - $E$ trace follows an exponential control function. When the electric field reaches a certain value (5 ±2 kV / cm), the current density begins to drop, showing a NDR characteristic. The resistance state is translated from a low-resistance state (LRS) to a high-resistance state (HRS). Then with the decrease of electric field, the $\sigma$ - $E$ trace follows another exponential control function. When the electric field reaches a low value, the $\sigma$ - $E$ curve becomes linear again. As the field changes under a negative polarity, a sudden jump in current density takes place before another NDR process. Contrarily, there is no obvious NDR effect for $Sr_3Co_2Z$ in dry and vacuum (0.1 Pa) environments with configuration i (curves 5 ~ 8), or in moist environments with configuration ii (curve 4). The results demonstrated in Fig. 1 indicate that the NDR effect is caused by water molecules in the ambient or moist environments and independent of gas type. For comparison, all of the published results for $Sr_3Co_2Z$ show a linear $\sigma$ - $E$ relation[24, 25, 32, 33]. The difference could be attributed to the lower electric fields applied in them than in this work.

Otherwise, a 'coercive field' of 5.3 kV / cm is obtained from $P$ - $E$ curve shown in Fig. 1. It corresponds well with the certain value (5 ±2 kV / cm) related to the NDR effect. However, the disappearance of NDR effect in dry and vacuum environments indicates that the $P$ - $E$ loop is independent of ferroelectric domain reversal. The loop is derived from the leakage current, just like that in banana skins[1].

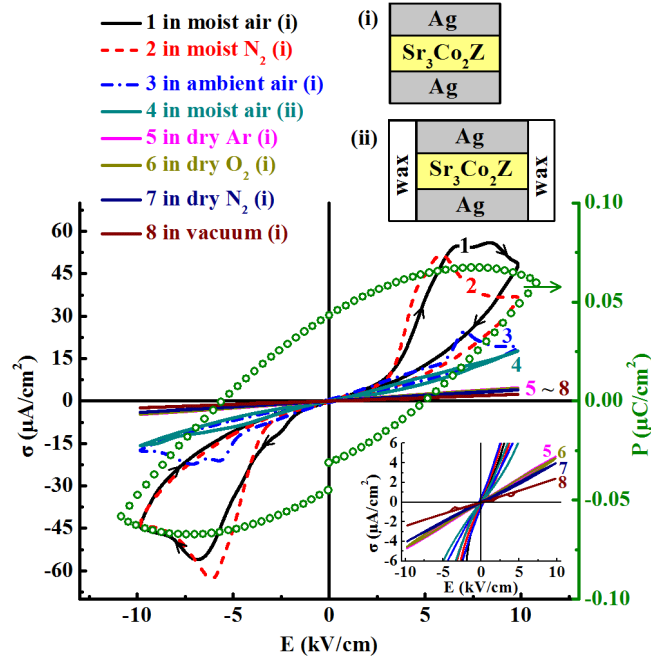

FIG. 1. Current density ($\sigma$) and polarization ($P$) as a function of electric field ($E$) for $Sr_3Co_2Z$ in various environment. The $\sigma$ - $E$ curves were tested with configuration i or ii (curves 4, configuration ii; the other curves, configuration i; $U_{max}$ = 840 V). The inset is the enlargement of curves 5 ~ 8 from -6 to 6 µA /cm$^2$. The $P$ - $E$ chart was tested in ambient air at 100 Hz with configuration i.

## B. Impedance spectroscopy

The impedance spectroscopy was tested for $Sr_3Co_2Z$ in dry and moist environments (Fig. 2) for a further investigation. The impedance in dry air is higher than that in moist air. The Cole-Cole plot can be well fitted by the presumed equivalent circuits. The values for the equivalent elements are listed in Table 1.

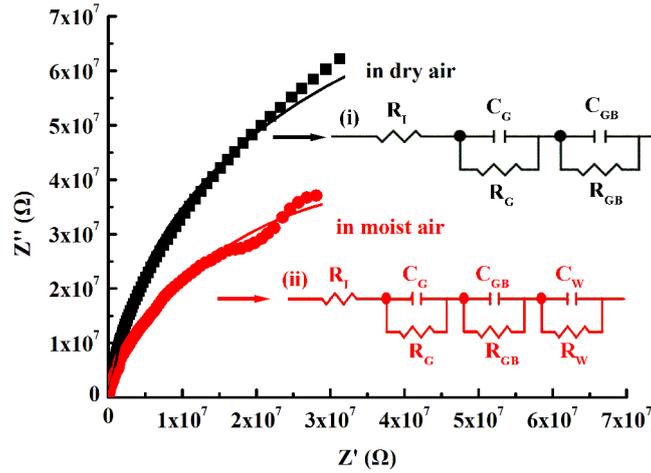

FIG. 2. Impedance spectroscopy for $Sr_3Co_2Z$ in dry and moist air. The dots are the raw data and the lines are the fitted results. The insets are the equivalent circuits in dry air (i) and in moist air (ii), respectively. $R_I$: interface contact resistance between the electrode and $Sr_3Co_2Z$; $C_G$, $C_{GB}$ and $C_W$: capacitance of grain, grain boundary and water film, respectively; $R_G$, $R_{GB}$ and $R_W$: resistance of grain, grain boundary and water film, respectively.

TABLE 1. The values of equivalent elements for $Sr_3Co_2Z$ in dry air and moist air.

| Environments | $R_I$ (Ω) | $C_G$ (pF) | $R_G$ (MΩ) | $C_{GB}$ (pF) | $R_{GB}$ (MΩ) | $C_W$ (pF) | $R_W$ (MΩ) |
|---|---|---|---|---|---|---|---|
| Dry air | 363.3 ±0.8 | 62.3 ±0.2 | 1.0 ±0.2 | 18.5 ±0.3 | 142.9 ±0.6 | - | - |
| Moist air | 455 ±72 | 44.2 ±4.2 | 6.0 ±0.5 | 32.6 ±2.6 | 66.8 ±7.7 | 61.1 ±1.7 | 0.22 ±0.01 |

The resistance of grain boundary ($R_{GB}$) is much larger than the others, indicating that the conduction current could be limited by the electrical properties of the grain boundaries[34]. The resistivity derived from the impedance data in dry air for $Sr_3Co_2Z$ is $1.85 \times 10^9$ Ω·cm, which corresponds well with that calculated from $\sigma$ - $E$ curve in vacuum ($4.09 \times 10^9$ Ω cm). As in moist air, the voltage on water film is about 1.265 V where the current begins to decrease, larger than the water-decomposition voltage of 1.229 V [5 kV / cm is corresponding to 420 V, while $420 \times 0.22 / (6.0 + 66.8 + 0.22) \approx 1.265$].

## C. Discussion

When the water molecules are absorbed on the lateral surface of a sample that is not covered by electrode metal (see configuration i in Fig. 1), they will diffuse along the channel between the electrode (Ag) and the material ($Sr_3Co_2Z$). The water molecules could also diffuse along the grain boundary resulting in the decrease of resistivity (Table 1). The voltage for water decomposition is 1.229 V under the standard condition (298 K and 1 atm) (Eqs. 1 ~ 3), above which the water could be decomposed into $H_2$ and $O_2$. This corresponds well with the voltage on water where the current begins to decrease (~ 1.265 V). Thus, the NDR effect could be induced by the decomposition of water.

$$O_2 + 2H_2O + 4e^- = 4OH^- \qquad E^\theta = 0.401V \tag{1}$$

$$2H_2O + 2e^- = H_2 + 2OH^- \qquad E^\theta = -0.828V \tag{2}$$

$$2H_2O = 2H_2 + O_2 \qquad E^\theta = 0.401 + 0.828 = 1.229V \tag{3}$$

The $\sigma$ - $E$ curve of $Sr_3Co_2Z$ in moist or ambient condition (curves 1 ~ 3 in Fig. 1) can be divided into five regions (Fig. 3, take curve 1 in Fig. 1 for example).

**Region I** (ohmic conduction region). In Region I (0 ≤ $E$ ≤ 2.5, kV / cm), the $\sigma$ - $E$ shows a linear feature. The pronounced grain boundary barriers make a predominant contribution to the series resistance. Conduction through the grain boundaries at low voltages is probably due to thermally activated excitation of the charge carriers and may consequently be described by Eq. 4 ($\rho_M$ is the resistivity of material)[35].

$$\sigma = \frac{E}{\rho_M} \tag{4}$$

**Region II** (tunneling region). In Region II ($2.5 < E \leq 6.5$, kV / cm), $\sigma$ increases exponentially with the electric field, which indicates an electron tunneling effect[8, 35]. The increased electric field leads the grain boundary barriers to be overcome. Correspondingly, the current increases dramatically, giving rise to a dielectric-breakdown and a tunneling process. The $\sigma$ (A / m$^2$) could be well fitted by Eq. 5[8, 36, 37]. $U$ is the voltage on each grain boundary (V), $\varphi_0$ is the barrier height (V), $d$ is the tunneling distance, *i.e.*, thickness of the grain boundary (m), $h$ is the Planck's constant (J s), $\alpha$ is a unitless adjustable parameter used in fitting, which is related to the barrier shape and the effective mass of an electron[8, 37], $m$ is the rest mass of an electron (kg) and $e$ is the elementary charge (C).

The average grain size of Sr$_3$Co$_2$Z derived from SEM is about 5 μm (not shown here) which is comparable with the previously reported[38]. For simplicity, a layered model is used for simulating the material (the inset on top left in Fig. 3). For a sample with a height of 0.84 mm, there are about 168 layers of grain (The thickness of grain boundary could be neglected compared with the size of grain.). Another assumption is that all of the voltage is applied on the grain boundaries because of their high resistance. Thus, there will be about 3.25 V on each grain boundary when the voltage reaches 546 V (6.5 kV / cm). A reasonable fitting result is obtained with a barrier height of 10.5 V, a grain boundary thickness of 1.2 nm and $\alpha = 0.89$ (Fig. 3).

$$\sigma = \frac{e}{2\pi h d^2}\left\{\left(e\varphi_0 - \frac{eU}{2}\right) \times \exp\left[-\frac{4\pi\alpha d}{h}(2m)^{\frac{1}{2}}\left(e\varphi_0 - \frac{eU}{2}\right)^{\frac{1}{2}}\right]\right. \\ \left. -\left(e\varphi_0 + \frac{eU}{2}\right) \times \exp\left[-\frac{4\pi\alpha d}{h}(2m)^{\frac{1}{2}}\left(e\varphi_0 + \frac{eU}{2}\right)^{\frac{1}{2}}\right]\right\} \tag{5}$$

**Region III** (NDR region). In Region III ($6.5 < E \leq 10$, kV / cm), $\sigma$ decreases with electric-field increasing, resulting in a NDR effect. When electric field increases in this region, the absorbed water decomposes continuously. The barrier height ($\varphi_0$) increases correspondingly and causes the drop of current.

**Region IV** (tunneling region). In Region IV ($2.5 > E \geq 10$, kV / cm), $\sigma$ - $E$ curve exhibits an exponential function as in Region II. The curve can also be well fitted by Eq. 5. During the fitting process, the thickness of grain boundary ($d$) is fixed while $\varphi_0$ and $\alpha$ are variable. Compared with the result in

Region II, the barrier height increases from 10.5 V to 17.2 V due to the decomposition of water molecules, leading to the smaller $\sigma$. Besides, water molecules are absorbed again gradually on the sample surface when the electric field is decreasing.

**Region V** (ohmic conduction region). In region V ($0 > E \geq 2.5$, kV / cm), a linear feature in $\sigma$ - $E$ curve and the conduction described by Eq. 4 are similar to those in Region I.

The above discussions are correlative to ambient and moist environments with configuration i (curves 1 ~ 3 in Fig. 1). While in dry and vacuum environments with configuration i (curves 5 ~ 8 in Fig. 1) or in moist environment with configuration ii (curve 4 in Fig. 1), there is little water molecules absorbed by the sample. The barrier height could not be overcome through the whole voltage range. As a result, $\sigma$ - $E$ curves show a linear feature which could be expressed by Eq. 4.

Thus, the NDR effect could be induced independently by the absorbed water and irrespective of intrinsic properties of materials / devices. The barrier height of grain boundaries is reduced by water molecules, giving rise to a dramatically increased current. Afterwards, with water decomposing, the barrier height of grain boundaries increases gradually, leading to a current decreasing, *i.e.*, the a NDR effect.

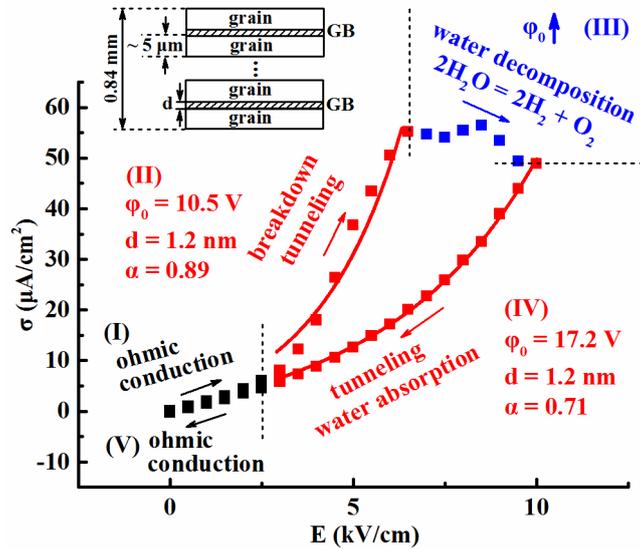

FIG. 3. Mechanism and the fitted results for the observed $\sigma$ - $E$ curve 1 in Fig. 1. The dots are raw data and the lines are the fitted results according to Eq. 5. The inset on top left is the simplified model for simulating. GB represents the grain boundary.

The $\sigma$ - $E$ curves were also tested for $TiO_2$, $Al_2O_3$, glass and banana skins (Fig. 4a, 4b, 4c and 4d). In moist or ambient air, the NDR effect was observed in all of the samples. But there was no NDR effect found when samples were put in dry air (except banana skins which were not dried). The results support the standpoint that a NDR effect could be induced by an extrinsic environment.

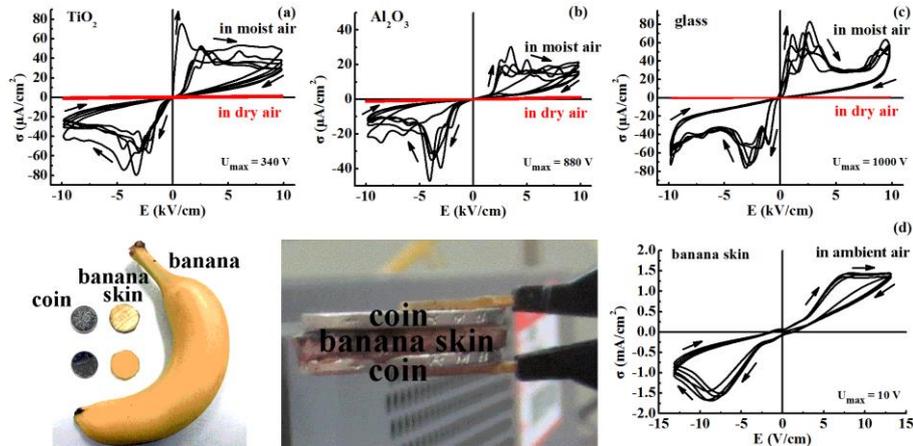

FIG. 4. Current density ($\sigma$) as a function of electric field ($E$) (5 loops) for $TiO_2$ (a), $Al_2O_3$ (b), glass (c) and banana skins (d). Banana skins were tested by fixing them between two coins (as electrodes).

## IV. CONCLUSION

If you observe close hysteresis loops with apparent dips in a current-voltage ($I$ - $V$) curve, it may be derived from another 'banana'. Bananas are not negative differential resistors, and it is easy to be misled by the loops. The NDR effect could only come from the absorbed water on a solid, whatever the solid is a real NDR material or not.

## ACKNOWLEDGEMENTS

The authors gratefully acknowledge Ms. W. Cai, Ms. X.-H. Yu and Mr. H.-X. Gu for their previous work in our group, Prof. Z.-G. Sun and Prof. J.-F. Wang in WUT, Prof. J. Xu in WIT and Prof. Z.-C. Xia in HUST for the helpful discussion. This work was supported by grants from State Key Laboratory of Advanced Technology for Materials Synthesis and Processing (WUT, China) (Grant Nos. 2010-PY-4, 2013-KF-2, 2014-KF-6), the Open Research Foundation of Key Laboratory of Nondestructive Testing (NHU, China) (Grant No. Zd201329002), the National Natural Science Foundation of China (Grant No. 51172049) and the Special Prophase Project on National Basic Research Program of China (Grant No. 2012CB722804).